# Quantum Mirrors Via SPDC-Crossing Symmetric Phenomena and A New Geometric Optics

D. B. Ion, P. Constantin and M.L.D. Ion

National Institute for Physics and Nuclear Engineering, Horia Hulubei IFIN-HH,
P.O.Box MG-6, Romania

**Abstract**: In this paper, the crossing symmetry of the photons, from spontaneous parametric down conversion (SPDC) phenomena in nonlinear crystals, is used for the development of a new geometric optics based on quantum kinematic correlations. On this basis, a new interpretation of the recent results on the two photon entangled experiment is obtained. A new field of applications is suggested.

**PACS**: 42.50. Tv; 42.50. Ar; 42.50. Kb; 03.65. Bz.

**Introduction**

It is well known that optical phase conjugation (OPC) (see e.g. Refs.[1-5] is a technique that incorporate a large variety of nonlinear effects to precisely reverse both the direction of propagation and the overall phase factor for each plane wave in an arbitrary beam of light. The spontaneous parametric down conversion (SPDC) is a nonlinear optical process [6] in which a laser pump (p) beam incident on a nonlinear crystal leads to the emission of correlated pair of photons called signal (s) and idler (i). In this process energy and momentum of photons are conserved. In this paper a new geometric optics called quantum SPDC-geometric optics is systematically developed by using kinematical correlations of the pump, signal and idler photons from the SPDC processes. We find that the new geometric SPDC-optics reproduce very well the results of the recent experiments [7-9] on the biphoton optics.

**2. Geometric Optics of Photons from SPDC Phenomena**

Recently [7-9], the SPDC process allowed to demonstrate two-photon "ghost" imaging and two-photon "ghost" interference-diffraction patterns, as well as other new phenomena from the optics of biphotons. In a recent paper [10] new interpretations of these phenomena is obtained in terms of the crossing symmetric SPDC processes.

2.1. *Crossing symmetric SPDC-processes as phase conjugation replicas*. If the S-matrix crossing symmetry [11] of the electromagnetic interaction in the spontaneous parametric down conversion (SPDC) crystals is taken into account, then the existence of the direct SPDC process

$$p \rightarrow s + i \quad (1)$$

will imply the existence of the following crossing symmetric processes

$$p + \bar{s} \rightarrow i \quad (2)$$

$$p + \bar{i} \rightarrow s \quad (3)$$

as real processes which can be described by the same transition amplitude. Here, by $\bar{s}$ and $\bar{i}$ we denoted the time reversed photons relative to the original photons s and i, respectively.

In fact, the SPDC-effects (1)-(3) can be identified as being directly connected with the $\chi^{(2)}$ − second-order nonlinear effects called in general three wave mixing (see Ref.[1,12]). So, the process (1) is just the inverse of second-harmonic generation, while, the effects (2)-(3) corresponds to the emission of optical phase conjugated replicas in the presence of pump laser via three wave mixing. Indeed, as was first pointed out by Yariv [2] and proved experimentally by Avizonis et al. [3], one can use the method of three-wave mixing to generate phase-conjugate replicas of any optical beam. This scheme exploits the second order optical nonlinearity in a crystal lacking inversion symmetry. In such crystals, the presence of input pump field (p) and signal wave (s)

$$E_p = \frac{1}{2} E_p(\omega_p) \exp\left(i\left[\omega_p t - \vec{k}_p \cdot \vec{r}\right]\right) + c.c. \quad (4)$$

$$E_s = \frac{1}{2} E_s(\omega_s) \exp\left(i\left[\omega_s t - \vec{k}_s \cdot \vec{r}\right]\right) + c.c.$$

induces in the medium a nonlinear optical polarization (see Eqs. (26)-(27) in Pepper and Yariv Ref.[1]) which is:

$$P_i = \chi_{ijk}^{(2)} E_{pj}(\omega_p) E_{sk}^*(\omega_s) \exp\left(i\left[(\omega_p - \omega_s)t - (\vec{k}_p - \vec{k}_s)\cdot\vec{r}\right]\right) + c.c. \tag{5}$$

where $\chi_{ijk}^{(2)}$ is the susceptibility of rank two tensor components of the crystal. Consequently, such polarization, acting as a source in the wave equation will radiate a new wave $E_i(\omega_i)$ at frequency: $\omega_i = \omega_p - \omega_s$ with an amplitude proportional to $E_i^*(\omega_i)$, i.e., to the complex conjugate of the spatial amplitude of the low-frequency probe wave at $\omega_{pw}$. Moreover, it was shown [1,2] that a necessary condition for a phase-coherent cumulative buildup of conjugate-field radiation at $\omega_i = \omega_p - \omega_s$ is that the wave vector $\vec{k}_i$ at this new frequency must be equal to $\vec{k}_i = \vec{k}_p - \vec{k}_s$, i.e., we have

$$\omega_i = \omega_p - \omega_s, \quad \vec{k}_i = \vec{k}_p - \vec{k}_s \tag{6}$$

This condition can be satisfied along only one direction in crystal by using the optical anisotropy to compensate for the (linear) refractive index dispersion. Hence, the optical phase conjugation by three-wave mixing (OPC-TWM) [1,2] help us to obtain a complete proof of the existence of the crossing reactions (2)-(3) as real processes which take place in SPDC crystals, when the phase matching conditions (6) are fulfilled. So, if $(\omega_p, \vec{k}_p), (\omega_s, \vec{k}_s), (\omega_i, \vec{k}_i)$ are the energy and momentum for the pump (p), signal (s) and idler (i) photons, respectively, then, the phase matching conditions (6) can be recognized as the energy-momentum conservation relations $(\eta = c = 1)$ for all SPDC processes (1)-(3). In fact, a rigorous phenomenological crossing symmetric theory for the non degenerate TWM in nonlinear media, which include all three types of second-order susceptibility tensors with their Kleinsman's permutation symmetry, can be developed in similar way with that discussed in Refs. [4,5].

2.2. *Quantum mirrors via SPDC phenomena.* The main purpose here is to obtain an answer to the basic question: what is a quantum mirror and what does it do?.
**D.1** QUANTUM MIRROR. By definition a quantum mirror (QM) is a combination of standard devices (e.g., usual lenses, usual mirrors, lasers, etc.) with a nonlinear crystal by which one involves the use of a variety of quantum phenomena to exactly transform not only the direction of propagation of a light beam but also their polarization characteristics.

**D.2** SPDC-QUANTUM\ MIRROR (QM). A quantum mirrors is called SPDC-QM if is based on the quantum SPDC phenomena (1)-(3) in order to transform signal photons, characterized by $(\omega_s, \vec{k}_s, \vec{e}_s, \mu_s)$, into idler photons with $(\omega_p - \omega_s, \vec{k}_p - \vec{k}_s, \vec{e}_s^*, -\mu_s) \equiv (\omega_i, \vec{k}_i, \vec{e}_i, \mu_i)$.

The high quality of the SPDC-QM is given by the following peculiar characteristics:
(i) *Coherence*: The SPDC-QM preserves high coherence between s-photons and i-photons;
(ii) *Distortion undoing*: The SPDC-QM corrects all the aberrations which occur in signal or idler beam path;
(iii) *Amplification*: A SPDC-QM amplifies the conjugated wave if some conditions are fulfilled.
Now, it is important to introduce the momentum projections, parallel and orthogonal to the pump-photon momentum, and to write momentum conservation from (6) as follows

$$k_p = k_s \cos\theta_{ps} + k_i \cos\theta_{pi} \tag{7}$$

$$k_s \sin\theta_{ps} = k_i \sin\theta_{ip} \tag{8}$$

where the angles $\theta_{pj}$, $j = s, i$, are the angles between momenta of the pump (p), signal (s) and idler (i) photons, respectively.

Clearly, a SPDC crystal illuminated by a high quality laser beam can acts as real quantum mirrors since by the crossing processes (2) (or (3)) a signal photon (or idler photon) is transformed in an idler photon $i_s$ (or signal photon $s_i$), respectively. The quantum mirrors can be "plane quantum mirrors"(P-QM) (see Fig.1) and "spherical quantum mirrors"(S-QM) (see fig.2) according with the character of incoming laser waves ( plane waves or spherical waves). In order to avoid many complications, in the following we will work only in the thin crystal approximation. Moreover, we do not consider here the so called optical aberrations.

**L.1** Law of thin plane SPDC-quantum mirror: Let BBO be a SPDC crystal illuminated uniform by a high quality laser pump. Let $Z_s$ and $Z_i$ be the distances shown in Fig.1 from the object point P to crystal (point A) and from crystal (point A) to image point I. Then, the system behaves as a plane mirror but satisfying the following important laws:

$$\frac{Z_i}{Z_s} = \frac{\omega_i}{\omega_s}, \quad M = \frac{\omega_s Z_i}{\omega_i Z_s} = 1 \tag{9}$$

where M is the linear magnification of the plane SPDC-quantum mirror.

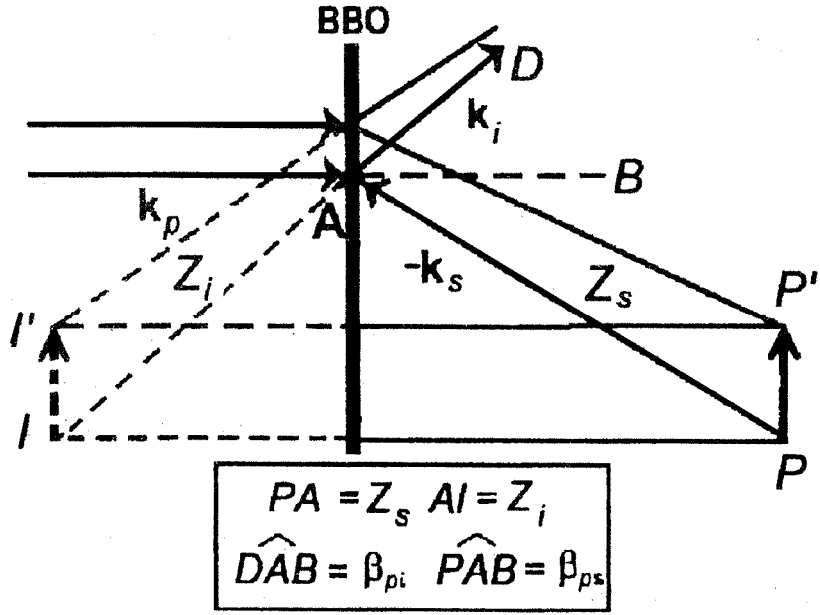

Fig. 1: The basic optical configuration of a plane SPDC-quantum mirror.

**L.2** *Law of thin spherical SPDC-quantum mirror:* Let QM be a SPDC crystal illuminated uniformly by a high quality laser pump (p), via a convergent lens (L) (with a focal distance f ) at distance d from crystal as in Fig.2a. Let, $Z_s$ and $Z_i$ be the distances PA and AI defined in Figs. 2a and 2b and let R=f-d be the distance AC. Then, the system behaves like a spherical mirror with the radius R in which the distances $Z_s$ and $Z_i$ and R, must satisfy the radial focal law

$$\frac{\omega_s}{Z_s} + \frac{\omega_i}{Z_i} = \frac{\omega_p}{R}\cos\beta, \quad M = \frac{\omega_s Z_i}{\omega_i Z_s}, \tag{10}$$

$$\cos\beta = \frac{\omega_s \cos\beta_{ps} + \omega_i \cos\beta_{pi}}{\omega_s + \omega_i} \leq 1$$

where M is the linear magnification of the spherical SPDC-quantum mirror. For the degenerate case: $\omega_s = \omega_i = \omega_p/2$, $\beta_{ps} = \beta_{ps} = \beta$ the relation (10) is just the law of the usual spherical mirrors

$$\frac{1}{Z_s} + \frac{1}{Z_i} = \frac{2}{R}\cos\beta, \quad M_0 = \frac{Z_i}{Z_s} \tag{11}$$

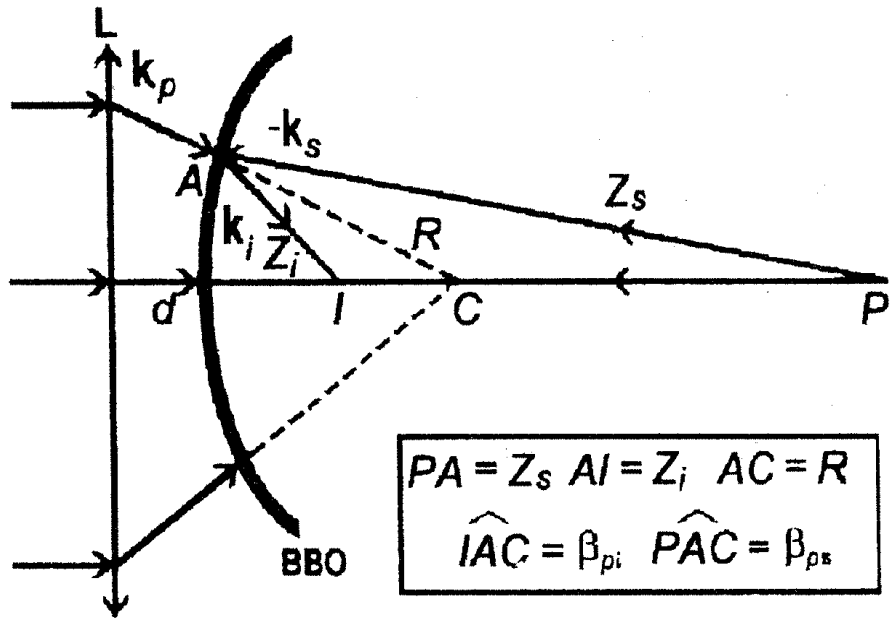

**Fig. 2a**: The basic optical configuration of a spherical SPDC-quantum mirror.

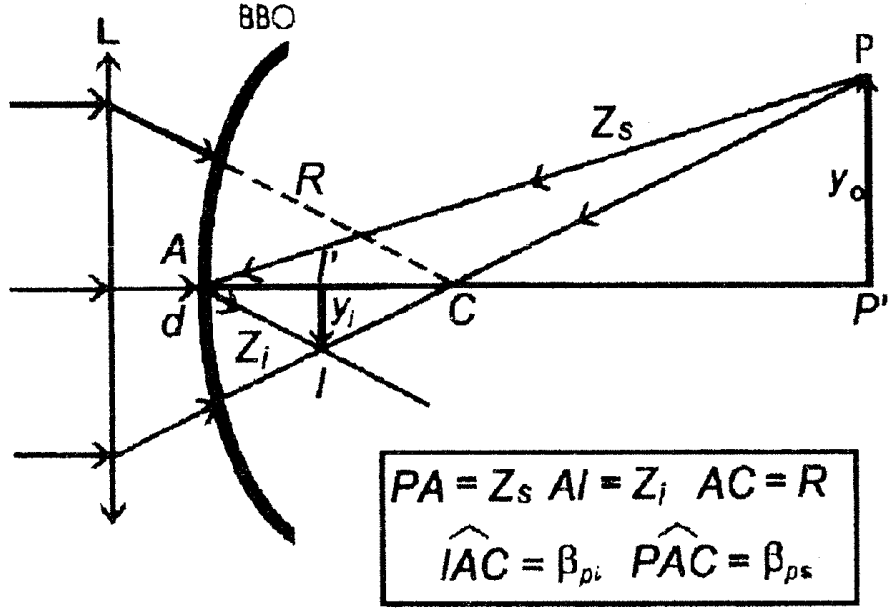

**Fig. 2b**: The basic optical configuration for a proof of magnification factor for a spherical SPDC-quantum mirror.

Proof: Here we prove only the law (10) starting from the basic optical configuration of SPDC-quantum mirrors shown in Fig.1a. Therefore, a photon emitted from the point P is propagating in direction of the SPDC-quantum mirror, attaining the crystal in the point A where with the pump photon is transformed in an idler photon via the crossing SPDC process (2) or, equivalently, as phase conjugated [5-7] replica of the signal photons in presence of the pump photons. From the point A the resulted idler photon is propagating in the direction of the point I. The main idea of the proof of the relation (10) is based on the quantum correlations (7)-(8) and observation that ( $\Delta \equiv$ triangle ):

$$Area(\Delta PAI) = Area(\Delta PAC) + Area(\Delta ACI) \tag{12}$$

Indeed, by direct calculus we have:

$$Area(\Delta PAC) = \frac{1}{2} Z_s R \sin\beta_{ps} \tag{13}$$

$$Area(\Delta ACI) = \frac{1}{2} Z_i R \sin\beta_{pi} \tag{14}$$

$$Area(\Delta PAI) = \frac{1}{2} Z_s Z_i \sin(\beta_{ps} + \beta_{pi}) \tag{15}$$

where the angles $\beta_{ps}$, $\beta_{pi}$ are the exiting angles from the crystal shown in Fig.(2a). It can also be seen from the transverse components (8) taken in conjunction with the Snell's law upon exiting the crystal that

$$\omega_s \sin\beta_{ps} = \omega_i \sin\beta_{pi} \tag{16}$$

Hence, as a consequence of the relations (16), we have

$$\sin(\beta_{ps}+\beta_{pi}) = \sin\beta_{ps}\cos\beta_{pi}+\sin\beta_{pi}\cos\beta_{ps} = \qquad(17)$$

$$\sin\beta_{ps}[\cos\beta_{pi}+\frac{\omega_s}{\omega_i}\cos\beta_{ps}] = \frac{[\omega_i\cos\beta_{pi}+\omega_s\cos\beta_{ps}]}{\omega_i}\sin\beta_{ps}$$

So, by using (13)-(16), from (12), by dividing with $Z_sZ_iR\sin\beta_{ps}/2$ we get the result (10). A proof of the magnification factor can be obtained on the basis of geometric optical configuration from Fig. 2b. Hence, the magnification factor is

$$M = \frac{y_i}{y_o} = \frac{Z_i\sin\beta_{ps}}{Z_s\sin\beta_{pi}} = \frac{\omega_s Z_i}{\omega_i Z_s} \qquad \text{(using (16))} \qquad (18)$$

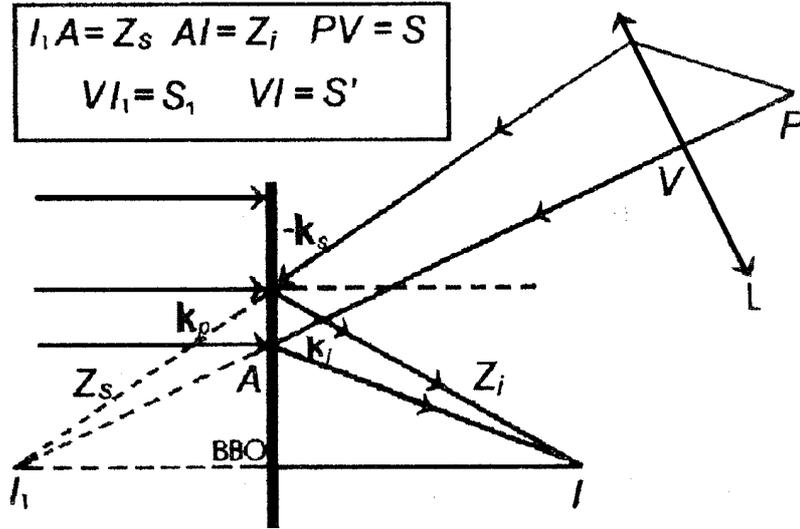

**Fig. 3a:** The basic optical configuration for usual lens combined with a plane SPDC-quantum mirror.

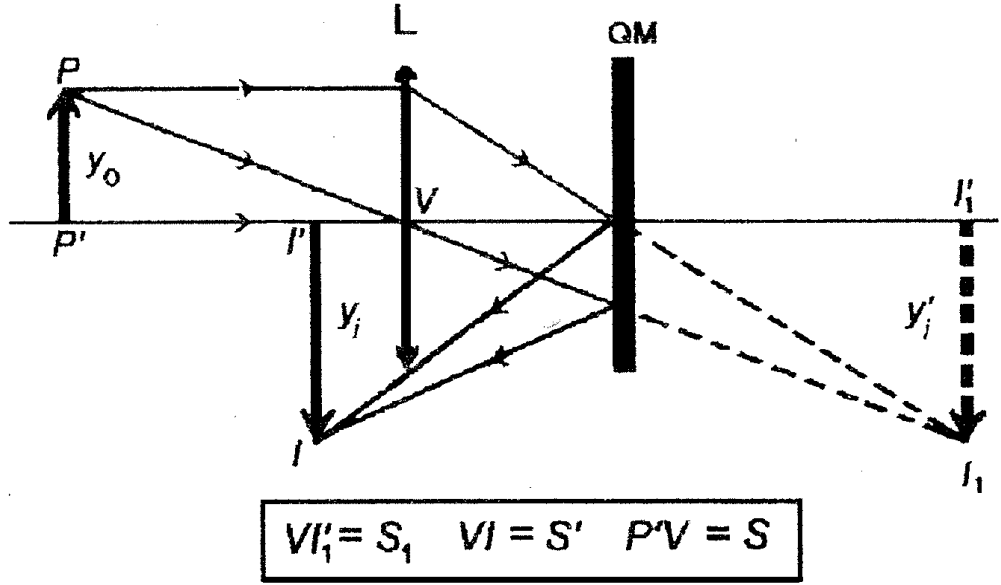

**Fig. 3b**: The basic optical configuration for a proof of magnification factor for a usual lens combined with a plane SPDC-quantum mirror.

Therefore, the key of the law (10) for the thin spherical SPDC-quantum mirrors is given by the quantum kinematical correlations (7)-(8) of the SPDC process (1).
Now, it is important to note that other kinds of quantum mirror, such as parabolically quantum mirrors, hyperbolical quantum mirrors, etc., can be obtained by the appropriate transformations of the incident pump wave front, by using different types of lenses before the SPDC crystal.

2.3. *Plane SPDC-QM combined with thin lens*. The basic optical geometric configurations of a plane SPDC-QM combined with thin lens is presented in Figs. 3a and 3b. In this case the system in behaves as in usual geometric optics, but with some modifications in the non degenerate case introduced by the presence of the plane SPDC-quantum mirror. The remarkable law in this case is as follows.

**L.3** *Law of the thin lens combined with a plane SPDC-QM*: The distances S (lens-object), S'(lens-crystal-image plane), $D_{CI}$ (crystal-image plane) and f (focal distance of lens), satisfy the following thin lens equation

$$\frac{1}{S} + \frac{1}{S' + \left(\frac{\omega_s}{\omega_i} - 1\right)D_{CI}} = \frac{1}{f} \tag{19}$$

The SPDC-QM system in this case has the magnification M given by

$$M = \frac{S' + \left(\frac{\omega_s}{\omega_i} - 1\right)D_{CI}}{S} = M_o + \left(\frac{\omega_s}{\omega_i} - 1\right)\frac{D_{CI}}{S} \tag{20}$$

In degenerate case $\omega_s = \omega_i = \omega_p/2$ we obtain the usual Gauss law for thin lens with the magnification $M_o = \frac{S'}{S}$.

Proof: The proof of the predictions (19)-(20) can be obtained by using the basic geometric optical configuration presented in Fig. 3a. Hence, the image of the object P in the thin lens placed between the crystal and object is located according to the Gauss law

$$\frac{1}{S} + \frac{1}{S_1} = \frac{1}{f} \tag{21}$$

where $S_1$ is the distance from lens to image $I_1$. Now the final image I of the image $I_1$ in the plane SPDC-QM is located according to the law (9). Consequently, if d is the lens-crystal distance then we have

$$S_1 = S' + (Z_s - Z_i) = S' + \left(\frac{\omega_s}{\omega_i} - 1\right)D_{CI} \tag{22}$$

since $S_1 = d + Z_s$, $S' = d + Z_i$ and $D_{CI}$ is the crystal-image distance. A proof a the magnification factor can be obtained on the basis of geometric optical configuration from Fig. 3b. Hence, the magnification factor is

$$M = \frac{y_i}{y_o} = \frac{y_i}{y'_i}\frac{y'_i}{y_o} = \frac{y'_i}{y_o} \tag{23}$$

since the plane SPDC-QM has the magnification $\frac{y_i}{y'_i} = 1$. Obviously, from $\Delta PP'V \_ \Delta I_1 I_1'V$ we get $\frac{y'_i}{y_o} = \frac{S'}{S}$ and then, with (22) we obtain the magnification (20).

**L.4** *Law of thin lens + plane SPDC-QM with the null crystal-lens distance*

$$\frac{1}{S} + \frac{1}{\frac{\omega_s}{\omega_i}S'} = \frac{1}{f}, \quad M = \frac{\omega_s}{\omega_i}\frac{S'}{S} \tag{24}$$

Proof: Here we note that (L.4) is the particular case of (L.3) with d=0 for which we get $S_1 = Z_s$ and $S' = Z_i$ Then from (9) and (21) we obtain (24).

## 3. Experimental Tests for SPDC Quantum\ Mirrors

3.1 *Experimental tests for the plane SPDC-quantum mirror combined with thin lens*. For an experimental test of the Gauss like law of the thin lens combined with a plane SPDC-QM we propose an experiment based on a detailed setup presented in Fig. 4 and in the optical geometric configuration shown in Fig. 3b. Then, we predict that the image I of the object P (illuminated by a high quality signal laser SL with $s(\omega_s, \vec{k}_s, \vec{e}_s, \mu_s)$) will be observed in the idler beam with $(\omega_i, \vec{k}_i, \vec{e}_i, \mu_i) \equiv i(\omega_p - \omega_s, \vec{k}_p - \vec{k}_s, \vec{e}_s^*, -\mu_s)$, when distances lens-object (S), lens-crystal-image plane (S'), crystal-image plane ($D_{CI}$) and focal distance f of lens, satisfy thin lens+QM law (19). Moreover, if thin lens+QM law (19) is satisfied, the image I of that object P can be observed even when instead of the signal source SL we put a detector $D_s$.

This last statement is clearly confirmed recently, in the degenerate case $\omega_s = \omega_i = \omega_p/2$, by a remarkable two-photon imaging experiment [8]. Indeed, in this case an argon ion laser is used to pump a nonlinear BBO crystal $(\beta - BaB_2O_4)$ to produce pairs of orthogonally polarized photons (see Fig. 1 in Ref. [8] for detailed experimental setup). After the separation of the signal and idler beams, an aperture (mask) placed in front of one of the detectors ($D_s$) is illuminated by the signal beam through a convex lens. The surprising result consists from the fact that an image of this aperture is observed in coincidence counting rate by scanning the other detector ($D_i$) in the transverse plane of the idler beam, even though both detectors single counting rates remain constants. For understanding the physics involved in their experiment, they presented an "equivalent" scheme ( in Fig. 3 in Ref. [8]) of the experimental setup. By comparison of their "scheme" with our optical configuration from Fig. 3b we can identify that the observed validity of the two-photon Gaussian thin-lens equation

$$\frac{1}{f} = \frac{1}{S} + \frac{1}{S'} \qquad (25)$$

as well as of the linear magnification

$$M = \frac{S'}{S} = 2 \qquad (26)$$

can be just explained by our results on the two-photon geometric law (19)-(20) of the thin lens combined with a plane SPDC-QM for the degenerate case $\omega_s = \omega_i = \omega_p/2$. Therefore, the general tests of the predictions (10)-(11), using a setup described in Fig. 4, are of great importance not only in measurements in presence of the signal laser LS (with and without coincidences between LS and idler detector $D_i$), but also in the measurements in which instead of the laser LS we put the a signal detector $D_s$ in coincidence with $D_i$.

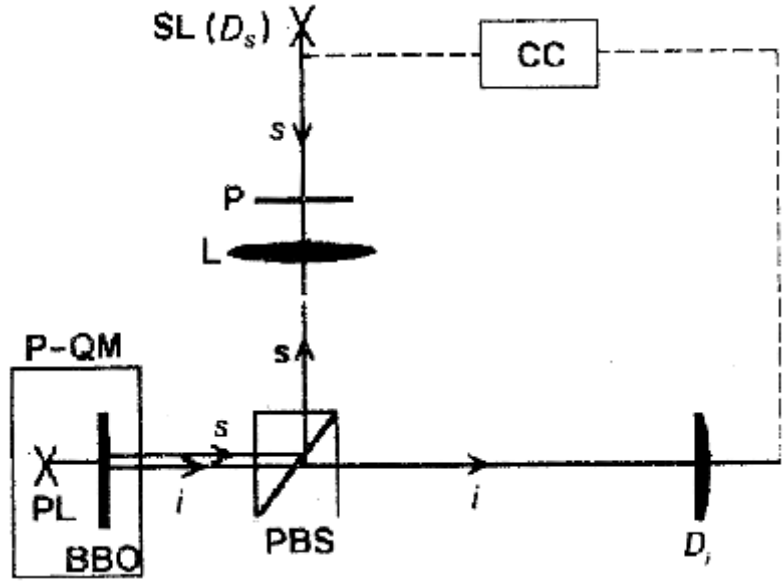

**Fig. 4**: The scheme of the experimental setup for a test of the two-photon geometric optics. The QM indicates the SPDC-quantum mirror, PBS is a polarization beam splitter, SL is a signal laser, P is an object, L a convergent lens, $D_i$ is an idler detector and CC is the coincidence circuit

3.2. *Experimental tests for the spherical SPDC-quantum mirrors.* Now, for an experimental test of the law of thin spherical SPDC-quantum mirror we propose an experiment based on a detailed setup presented in Fig. 4, but with the lens (L) removed and with a spherical SPDC-QM instead of a plane SPDC-QM. Then, we predict that the image I of the object P illuminated by a high quality signal laser SL will be observed in the idler beam, when the distances $Z_s, Z_i$ and R defined in Fig. 2a satisfy the law of the thin spherical SPDC-quantum mirror law (10) and (11). Moreover, if the relation (10) is satisfied, the image I of that object P will be observed in measurements in which instead of the signal laser SL we put a detector $D_s$ in coincidence with the idler detector $D_i$. This last prediction, as well as the magnification factor (11) was also evidenced with high accuracy recently by a two-photon geometric optics experiment [9], in both nondegenerate and degenerate cases.

3.3. *Other experiments related with SPDC-QM.* It is important to note that the experiments described above can also be realized in the variant in which the object P (see Fig.4) is replaced with any diffraction-interference apertures (as in Ref.

[7]). So, if thin lens+QM law (19) is satisfied, then, we expect that the diffraction-interference patterns can be observed in the detector $D_i$. Of course, the diffraction-interference patterns can be observed in coincidence measurements even when instead of the signal source SL we put a detector $D_s$. This is the case of the experiment [7]. Ribeiro et al [13] have obtained the following important result: the visibility of interference fringes produced by a signal beam transmitted through a double slit, can be controlled by aligning an auxiliary laser with the idler beam, with the same wavelength and varying its power. In this case, the degree of coherence of the source is varied directly by the inducing laser intensity without performing any measurements on the idler beam.

3.4. *Optical devices improvements via quantum mirrors*. Can the quantum mirrors produce a new revolution in the optical devices? For example, many telescopes (e. g., Newton, Gregory, etc.) are using parabolically mirrors as main image processing devices. If some usual mirrors will be replaced with appropriate quantum mirrors then we can obtain some major advantages. Some geometrical advantages (e.g., the magnifications and distances control by varying the pump wavelength, etc.) as well as signal processing advantages (e.g., the high resolution, the high fidelity and amplification of the incoming beam intensity, distortion undoing for the signal rays, coherence preserving, etc.) are expected to be obtained by using the quantum mirror instead of the usual mirrors.

**Discussions and Conclusions**
The main results and conclusions obtained in this paper can be summarized as follows:
(i) The class of SPDC-phenomena (1) is enriched by introducing the crossing symmetric processes (2)-(3) as real phenomena described just by the same transition amplitude as that of the original SPDC-process (1) and satisfying the same energy-momentum conservation law (6). The connection between the crossing symmetric reactions (2)-(3) and the well known optical phase conjugation phenomena is established. On this basis a new kind of devices, called quantum mirrors, is introduced (see (D.1)-(D.2)).
(ii) A new exotic geometric optics is developed on the basis of quantum mirrors. New experiments, for detailed investigation of the basic predictions of these new two photon geometric optics effects, are proposed.
(iii) New and complete interpretations of the experiments of the two-photon "ghost" image [8]-[9], as well as of the two-photon "ghost" interference diffraction experiments [7], are obtained.
(iv) Substantial improvements of the optical devices (e.g. telescopes, microscopes, etc.), are expected to be obtained by using quantum mirrors instead of the usual mirrors.
(This paper was published in Romanian Journal of Physics, Vol.**43**, Nos. 1-2, P. 3-14, Bucharest 1998)